\documentclass[pdflatex,sn-mathphys-ay]{sn-jnl}

\usepackage[utf8]{inputenc}
\usepackage[normalem]{ulem} 
\usepackage{microtype} 
\usepackage{amsmath} 
\usepackage{amssymb} 
\usepackage{algorithm} 
\usepackage{algpseudocode} 
\usepackage{url} 
\usepackage{tabularx} 
\usepackage{booktabs}
\usepackage{ragged2e}
\usepackage{xcolor}
\usepackage{lineno}

\newcommand{\deltasigned}[1]{%
  \begingroup
  \ifdim #1pt > 0pt
    \textcolor{blue}{#1}%
  \else\ifdim #1pt < 0pt
    \textcolor{red}{#1}%
  \else
    0%
  \fi\fi
  \endgroup
}

\definecolor{cool}{HTML}{0072B2}    
\definecolor{warm}{HTML}{D55E00}    

\definecolor{later}{HTML}{0f4a00}
\definecolor{earlier}{HTML}{7B3294}

\newcommand{\deltasignedday}[1]{%
  \begingroup
  \ifdim #1pt > 0pt \textcolor{later}{#1}%
  \else\ifdim #1pt < 0pt \textcolor{earlier}{#1}%
  \else 0\fi\fi
  \endgroup
}

\newcommand{\deltasignedtemp}[1]{%
  \begingroup
  \ifdim #1pt > 0pt \textcolor{red}{#1}%
  \else\ifdim #1pt < 0pt \textcolor{cool}{#1}%
  \else 0\fi\fi
  \endgroup
}

\let\orcidlogo\relax
\usepackage{orcidlink}

\begin{document}

\title{A multivariate index for estimating seasonal shifts from temperature curves}

\author*[1]{\fnm{Tinula} \sur{Kariyawasam} \orcidlink{0009-0005-8542-9918}}
\email{tinula.kariyawasam@student.rmit.edu.au\\Contact number: +61 423 031 004}

\author[1]{\fnm{Inge} \sur{Koch}}\email{inge.koch@rmit.edu.au}
\equalcont{These authors contributed equally to this work.}

\author[1]{\fnm{Gary} \sur{Bryant}}\email{gary.bryant@rmit.edu.au}
\equalcont{These authors contributed equally to this work.}

\author[1]{\fnm{Shuwen} \sur{Hu}}\email{shuwen.hu@rmit.edu.au}
\equalcont{These authors contributed equally to this work.}

\affil[1]{\orgdiv{School of Science}, \orgname{RMIT University}, \orgaddress{\street{124 La Trobe Street}, \city{Melbourne}, \postcode{3000}, \state{Victoria}, \country{Australia}}}

\abstract{Seasonal timing is often studied using fixed calendar seasons, temperature thresholds, or a fixed smoothing parameter applied across all years. These may be sensitive to changes in variability over time. To address this, we propose the \textbf{S}easonal \textbf{A}daptive \textbf{L}ocalisab\textbf{L}e \textbf{I}ndex (SALLI), a multivariate index, for characterising seasonality from daily temperature curves observed over complete seasons. SALLI uses Significant Zero Crossings of Derivatives (SiZer) to identify significant increases and decreases across a range of smoothing levels. It uses this information to select a smoothing level for each year. The method assumes one main warm peak and one cool trough per annual cycle. For each year, SALLI estimates the annual temperature curve and the midpoints of the four seasons and temperatures. These estimates describe changes in both seasonal timing and temperature. We illustrate SALLI using Australian maximum temperature data from 1965 to 2024. Comparisons of SALLI estimates over the first and last decades show not only a warming, but a shift to earlier seasons, with the midpoints of spring, summer, and winter occurring roughly a week earlier. All four seasonal midpoints become warmer, with the largest increases occurring in summer ($1.37~^\circ\mathrm{C}$) and spring ($1.22~^\circ\mathrm{C}$). Warming is evident across time, whereas the clearest shift in seasonal timing emerges after the 2000s. By estimating both timing and temperature at each seasonal midpoint, SALLI locates where changes occur on the annual temperature curve, providing information that is more robust and data-driven than fixed thresholds, fixed calendar seasons, or a single smoothing choice.}


\maketitle

\clearpage

\textbf{Highlights:}
\begin{itemize}
	\item \textbf{S}easonal \textbf{A}daptive \textbf{L}ocalisab\textbf{L}e \textbf{I}ndex (SALLI) is a new data-driven index for seasonal change.
	\item SALLI estimates annual seasonal timing and temperatures from daily temperature curves.
	\item An adaptive smoothing parameter is selected separately for each year.
	\item Australian maximum temperatures show an earlier seasonal cycle.
	\item The cooler half of the annual cycle is shortened by about six days.
\end{itemize}

\clearpage

\textbf{Graphical Abstract:}
\begin{center}
\includegraphics[width=\textwidth]{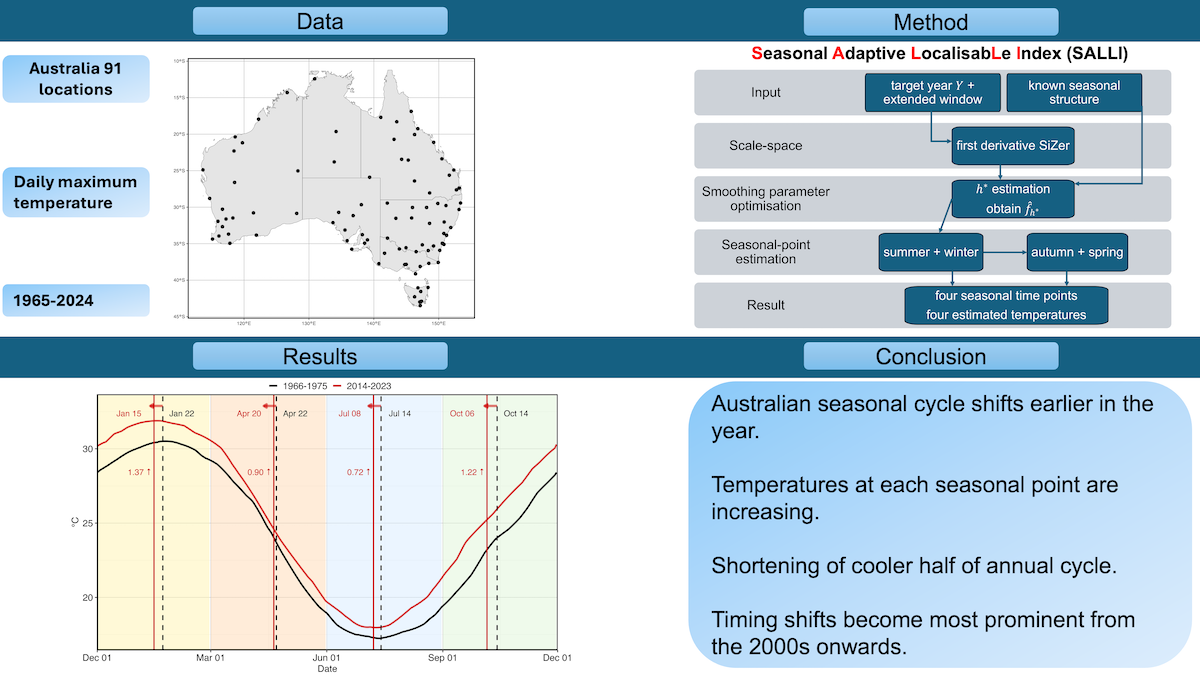}
\end{center}

\vspace{0.5em}

\noindent \textbf{Graphical Abstract Descriptions:}
This work presents the \textbf{S}easonal \textbf{A}daptive \textbf{L}ocalisab\textbf{L}e \textbf{I}ndex (SALLI), a new data-driven index for estimating seasonal timing shifts and temperature changes from daily temperature curves. Using Australian daily maximum temperatures from 91 locations over 1965-2024, the study demonstrates how seasonal change can be estimated directly from the annual temperature cycle rather than from fixed calendar seasons or temperature thresholds. SALLI analyses each year within an extended window and selects an adaptive smoothing parameter from a wide range of smoothing levels, choosing the minimum level that preserves the expected annual structure. The index then estimates four seasonal midpoints, summer, autumn, winter and spring, together with their fitted temperatures. The results show that recent Australian maximum-temperature cycles have shifted to earlier in the year, with higher temperatures at the seasonal midpoints and a shortening of the cooler half of the annual cycle. Overall, SALLI provides a robust curve-based framework for studying changes in both seasonal timing and temperature across any region in any continent.

\clearpage

\section{Introduction}
\label{sec:intro}

Global surface temperatures continue to break records, with 2024 estimated at \(1.52 \pm 0.13~^\circ\mathrm{C}\) above the 1850--1900 baseline \citep{forster2025}. Alongside this long-term warming, recent analyses report more frequent temperature swings and extremes across many regions, indicating an increase in climatic variability \citep{wu2025}. Regional environmental responses can reflect interactions between land-use changes and atmospheric conditions, as illustrated by long-term air-pollution patterns in Iran \citep{yousefi2025}. As climates warm and variability grows, seasonal timing can shift, summers may arrive earlier and last longer, winters may shorten, and transitional seasons can change character. Short-lived atmospheric events can produce marked changes in sea-surface temperature and water-column conditions, illustrating the importance of resolving short-term environmental variability \citep{lahijani2024}. These shifts matter for climate-sensitive sectors, including agriculture, health, tourism, energy systems and ecosystems, where planning depends on timely local information about when seasons begin and end, and how their lengths change over time \citep{ash2007, gossling2025, xu2025}. More broadly, climatic conditions interact with environmental processes across different timescales, from seasonal changes in lake-water properties \citep{mohammadi2023} to longer-term landscape evolution \citep{kavehfirouz2024}.

Seasons are commonly defined in two ways \citep{trenberth1983}. The astronomical definition provides a common reference based on the Earth's orbit, while the meteorological definition groups complete months to support consistent climate summaries. The astronomical definition is bounded by solstices and equinoxes (e.g., in the Southern Hemisphere, from the December solstice to the March equinox for summer). The more widely used meteorological or climatological definition partitions the year into four three-month blocks: December to February (summer), March to May (autumn), June to August (winter), and September to November (spring). These calendar seasons align neatly with monthly statistics, long observational records, and numerous applications, and historically, have approximated the hottest and coldest three-month periods reasonably well \citep{trenberth1983}. However, under contemporary warming and changing variability, fixed calendar or astronomical bounds can diverge from local thermal seasons. For decision-making, we increasingly require season definitions grounded in observed temperature behaviour at local and regional scales, with flexibility to track change.

A large body of work investigates seasonal timing changes using temperature thresholds. A common approach is to select fixed temperature values tailored to a country or region (e.g., winters below a specified threshold, summers above another) \citep{kitowski2019, lee2022, yiqi2023, fan2025}. At the same time, another uses percentile-based thresholds computed from the data \citep{dong2010, pena2015, allen2016, kutta2016, park2018, SwannOgge2020OutOfSeason, weller2020, wang2021, Scott_2026}. For example, \citet{lee2022} used fixed thresholds for Korea (e.g., winter \(< 5^\circ\)C, summer \(\ge 20^\circ\)C), and \citet{weller2020} identified Southern Hemisphere summers using the local 75th percentile of temperatures averaged over 60 years as a warm-season marker. Fixed and percentile-based thresholds provide interpretable estimates of the onset and duration of specified thermal conditions. Their suitability depends on the application, local climate, smoothing choices and stationary temperatures. However, assumptions of stationarity are hard to justify \citep{kariyawasam2025, castillo2023}.

Seeking greater flexibility, several studies employ clustering and other multivariate techniques to identify seasonal shifts from multiple variables \citep{kotsias2021, alghamdi2024}. These methods can uncover regional structure and suggest longer summers in some regions. However, they also introduce additional choices, including the number of clusters, feature scaling, and initialisation, which can complicate comparisons across heterogeneous climates. Another line of research avoids explicit start/end detection by focusing on the annual maxima and minima, combined with smoothing to infer 90-day seasons \citep{hekmatzadeh2020}. This improves identifiability of the seasonal features but still imposes a fixed season length and relies on a particular smoothing parameter choice. This focus on individual seasonal features can be extended to consider the annual temperature curve as a whole. Such a perspective includes examining human influence on the amplitude of the temperature cycle \citep{santer2018} and using functional data analysis to describe changes in the shape and timing of the mean annual temperature curve \citep{holtanova2026}. 

Across these approaches, daily temperatures are almost always smoothed before identifying seasons. The studies discussed above employ moving averages with varying window widths, low-degree polynomials, or harmonic fits. The smoothing window is usually fixed in advance; after such analyses, authors often note in discussion that the chosen window was too narrow or too broad, and that results were sensitive to this choice \citep{allen2016, kitowski2019}. To limit the influence of brief spikes, many studies impose persistence rules (for example, several consecutive days above a threshold). In practice, the key issue is how much to smooth; too little and day-to-day noise dominates, while too much obscures genuine shifts in onset, withdrawal, and inflection structure. A data-driven, robust treatment is therefore required.

Addressed here is how to determine the optimal amount of smoothing for estimating seasonal timing and fitted temperatures using consistent feature definitions across locations separately for each yearly cycle. A fixed smoothing level may retain short-term fluctuations in some series and flatten seasonal features in others, affecting both timing and temperature estimates.

To address these challenges of location-dependent season definitions and sensitivity to fixed smoothing choices, we propose the \textbf{S}easonal \textbf{A}daptive \textbf{L}ocalisab\textbf{L}e \textbf{I}ndex (SALLI). SALLI reframes seasonal timing as a set of points on the annual temperature curve, rather than as fixed start and end dates. The index uses the annual temperature cycle to select an optimal smoothing parameter for each year, with Significant Zero Crossings of Derivatives (SiZer; \citealp{sizer1999}) providing the flexibility of a continuous interval of smoothing parameters that show different features. Based on this range and the SiZer plot, we select an optimal smoothing parameter. We then use the resulting smoothed curve to estimate four seasonal points: summer, autumn, winter and spring midpoints and the estimated temperatures at each point. This allows SALLI to quantify both horizontal shifts in timing (earlier or later) and vertical changes in magnitude (cooler or warmer) in a single annual temperature curve. Here, we demonstrate SALLI using Australian daily maximum temperature data over 1965--2024. Since the focus of this paper is methodological, we apply the index at the national scale using an Australia-wide daily temperature curve. A companion study titled \emph{Estimating changes in temperature seasonality across hemispheres using the Seasonal Adaptive LocalisabLe Index (SALLI)} is currently in preparation. It applies SALLI to daily maximum and minimum temperature data from 15 countries across Europe, Asia, North America and Australia, extending the analysis to contrasting climatic settings and both hemispheres.

The remainder of the paper is organised as follows. Sect.~\ref{sec:methods} introduces the data and methods, including the SiZer framework and our choice criteria. Sect.~\ref{sec:results} presents methodological findings and examples. Sect.~\ref{sec:discussion} discusses the implications, limitations, and directions for broader application.

\section{Data and methods}
\label{sec:methods}

\subsection{Data}
\label{sec:data}

We use the Australian Climate Observations Reference Network – Surface Air Temperature (ACORN-SAT) dataset, curated by the Australian Bureau of Meteorology (BOM). ACORN-SAT provides homogenised daily maximum temperatures from high-quality stations distributed across Australia. Further information on design, homogenisation and quality control is available in \citet{trewin2020}. We analyse data from 1965 to 2024 inclusive.

\begin{figure}[ht]
 \centering
 \includegraphics[width=1\textwidth]{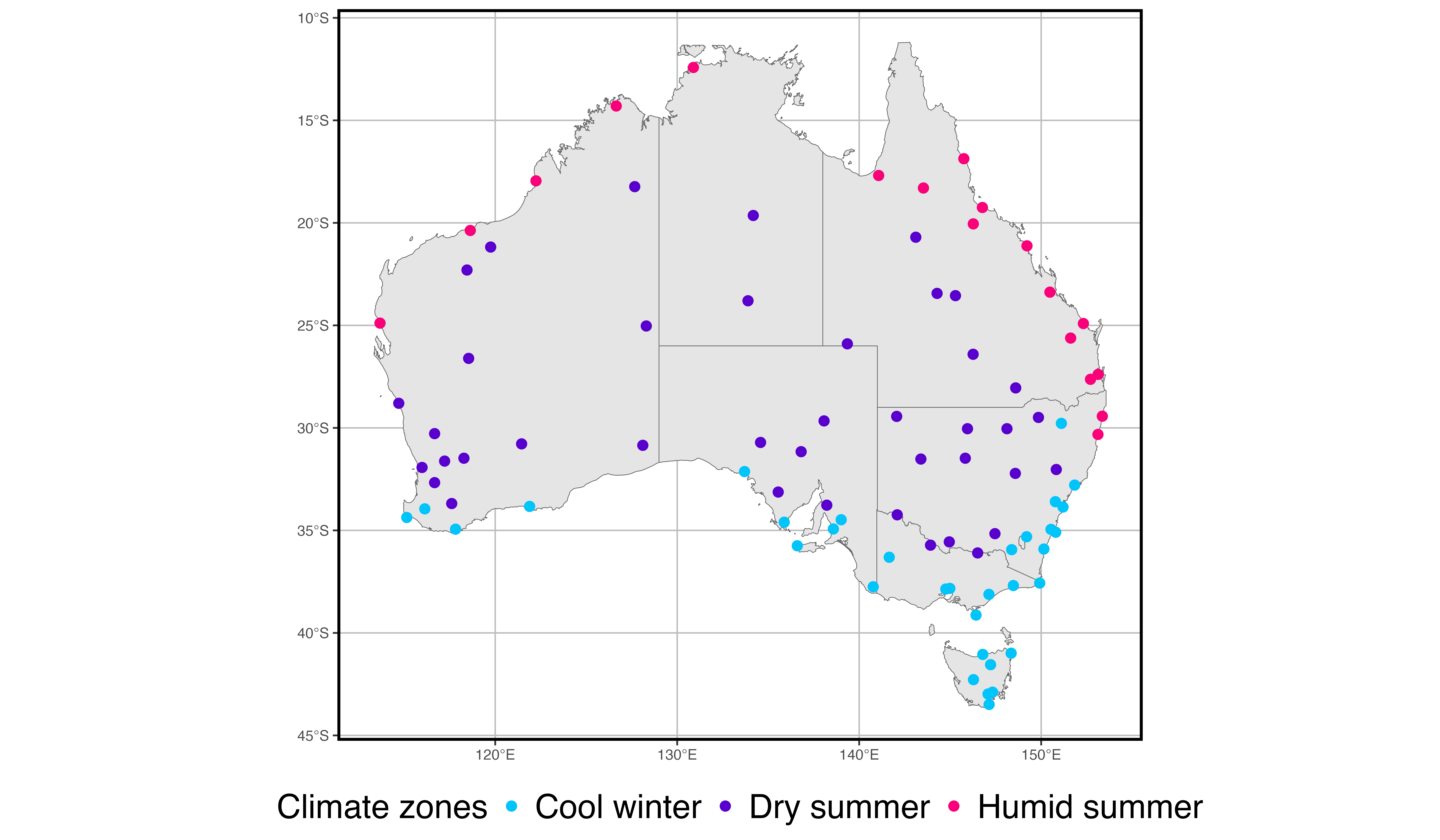}
 \caption{Locations of the 91 ACORN-SAT stations retained after completeness filtering (1965--2024). Colours distinguish the cool-winter, dry-summer and humid-summer groups based on the Bureau of Meteorology's climate classification.}
 \label{fig:Fig1}
 \end{figure}

To ensure data reliability, we retained stations with at least 95\% non-missing daily values over 1965--2024 and at most one year with $\ge 100$ missing daily values. This reduced the 110 available stations to 91. Leap days (29 February) are excluded to maintain a consistent 365-day calendar, a common practice in seasonal temperature analyses \citep{SwannOgge2020OutOfSeason}. Fig.~\ref{fig:Fig1} shows their spatial distribution. Based on the BOM's temperature and humidity climate classification \citep{bom_climateZones}, we also distinguish three broad climate-zone groups among the retained stations: cool winter, dry summer and humid summer. These groups are shown in Fig.~\ref{fig:Fig1} to provide context for the climatic diversity represented by the Australia-wide series.

Since this paper focuses on demonstrating the SALLI procedure at a broad spatial scale, several examples use Australia-wide daily median maximum temperature curves constructed from the 91 retained stations. Throughout the paper, when constructing an Australia-wide daily median curve over a set of years, we use the same two-stage median calculation. First, for each location and each day of the year, we take the median across the selected years. This gives one value per station for each day of the year. Second, for each day of the year, we take the median across the 91 station values, giving one Australia-wide temperature value per day. For single-year examples, the first step is unnecessary, and we take the median across the 91 station values for each date.

\subsection{From thresholds to features}\label{sec:motivation}

Daily temperatures are often smoothed before identifying seasonal structure. As discussed in Sect.~\ref{sec:intro}, existing approaches commonly use moving averages, low-degree polynomials, or harmonic fits, with the smoothing level chosen in advance. The main challenge is not the smoothing family itself, but the amount of smoothing to apply: too little smoothing leaves day-to-day noise, whereas too much smoothing can obscure genuine shifts in turning points and transition structure.

\begin{figure}[ht]
 \centering
 \includegraphics[width=1\textwidth]{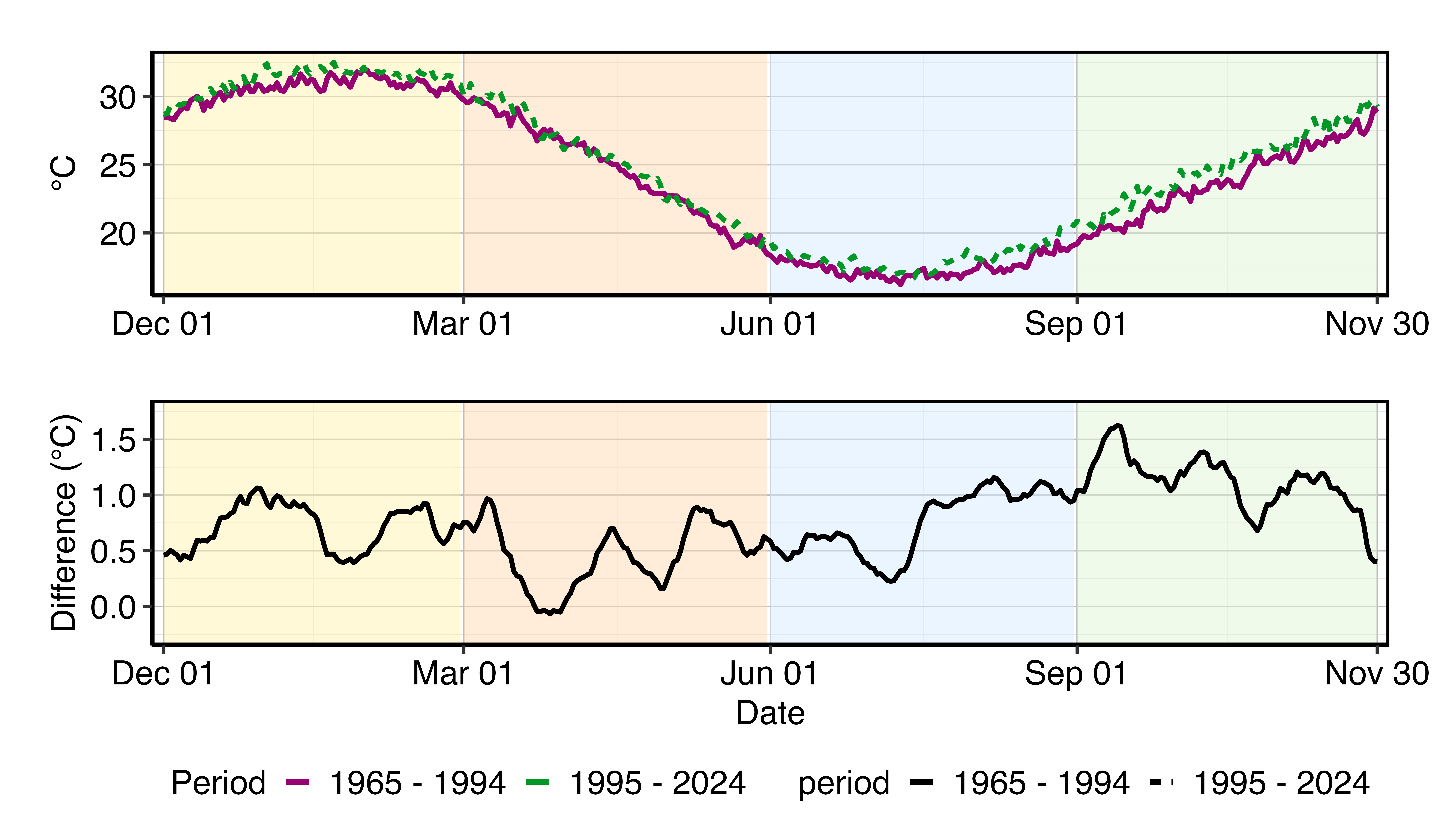}
\caption{Daily median maximum temperatures across 91 stations for 1965-1994 and 1995-2024 (top), and the corresponding day-of-year differences (1995--2024 minus 1965--1994; bottom). The \(x\)-axis starts on 1 December and runs through to the end of November. Shaded bands mark the conventional Southern Hemisphere three-month seasons: December--February (summer), March--May (autumn), June--August (winter), and September--November (spring).}
 \label{fig:Fig2}
 \end{figure}

Fig.~\ref{fig:Fig2} examines how warming in daily maximum temperature varies across the calendar year, rather than treating temperature change as a uniform shift. We split the data into two equal 30-year periods (1965-1994 and 1995-2024). For each period, we construct an Australia-wide daily median maximum temperature curve using the two-stage median calculation described in Sect.~\ref{sec:data}: first across years for each station and day of the year, and then across the 91 station-level values for each day. We then calculate the day-of-year temperature difference by subtracting the 1965--1994 curve from the 1995--2024 curve. This difference curve is smoothed using a 15-day centred moving average to highlight the broader structure. To align with Southern Hemisphere use, the $x$-axis begins in December and continues through the seasons to the end of November.

The lower panel shows that the difference between the later and earlier period is mostly between 0 and $1~^\circ\mathrm{C}$. However, there is a clear run of days from about 29 July to about 7 October where the smoothed difference exceeds $1~^\circ\mathrm{C}$, peaking around 7 September. These $71$ days extend from late July to early October, covering all of August and September. This pattern suggests that warming is not evenly distributed throughout the year. More broadly, Fig.~\ref{fig:Fig2} indicates that changes in the annual temperature cycle are not simply a uniform upward shift. The amount of warming varies across the year, suggesting that the timing and shape of the seasonal curve are also changing in ways that thresholds tied to fixed temperatures or percentiles may not capture.

These observations motivate a shift from threshold crossings to feature-based seasonality. Rather than assigning arbitrary start and end dates to seasons, we define points that can be extracted unambiguously from the annual temperature curve. The first two points describe the main turning regions of the cycle. The summer point represents the centre of the warm-season peak, while the winter point represents the centre of the cool-season trough. In practice, neither corresponds to a unique hottest day or coldest day. Instead, temperatures typically remain near their annual peak or trough for days or weeks, producing a broad peak or trough. Our aim is therefore to identify these warm and cool periods and characterise their timing by a representative centre point, rather than chase an arbitrary single-day extreme.

The other two points describe the transitions between the warm and cool parts of the curve. The autumn point marks the movement out of the warm season and the descent toward winter. The spring point marks the movement out of the cool season and the rise toward the next summer. This centres the analysis on the internal structure of the seasonal cycle, namely the warm-season peak, autumn transition, cool-season trough, and spring transition, rather than on nominal calendar season boundaries. Throughout the remainder of the paper, we refer to these four points as the summer, autumn, winter, and spring midpoints. The term midpoint is used for interpretation rather than imposed by definition. SALLI is not designed to find the midpoints of the conventional three-month calendar seasons. Instead, it estimates data-driven points on the annual temperature curve. In the Australian maximum-temperature series analysed here, these points often correspond approximately to the middle of the familiar seasonal periods, which makes the midpoint terminology useful for interpretation. Estimating these points robustly requires the right amount of smoothing to suppress day-to-day variability while preserving the main turning points and transition structure, a challenge we address in the following sections.

\subsection{Smoothing choices}\label{sec:smoothingLimits}

Daily temperatures are often smoothed before identifying seasonal structure. Across the research described in Sect.~\ref{sec:intro}, daily series are transformed using moving averages with fixed window widths, low-degree polynomials, or harmonic fits. In most cases, the smoothing window or model complexity is chosen and held fixed, with later discussion noting that the chosen setting was too short or too long, or that the results were sensitive to this choice. Persistence rules (for example, requiring several consecutive days above a threshold) are then added to reduce the influence of brief spikes. Choosing a smoothing level involves a trade-off; too little smoothing leaves day-to-day noise, whereas too much smoothing can obscure genuine shifts in turning points and onset structure and lead to under- or overestimation of temperatures at seasonal points.

The difficulty of achieving the ``right'' smoothing is well documented in the literature. \citet{kitowski2019} began with a 5-day moving average but found that their fixed thresholds were crossed too often, and so increased the window to 10 days. \citet{allen2016} experimented with 3, 5, 7 and 14-day windows before settling on a 10-day choice that they felt “allowed for seasonal fluctuations to exist yet minimised random cycling between seasons”. Different settings can be reasonable in isolation, but the inferred seasonal timing can still be sensitive to the specific smoothing level. The issue here is not the choice of smoothing family, for which many good options exist, but the assumption that a single, fixed degree of smoothing is usually sufficient across many years, stations, and climates.

\begin{figure}[ht]
 \centering
 \includegraphics[width=1\textwidth]{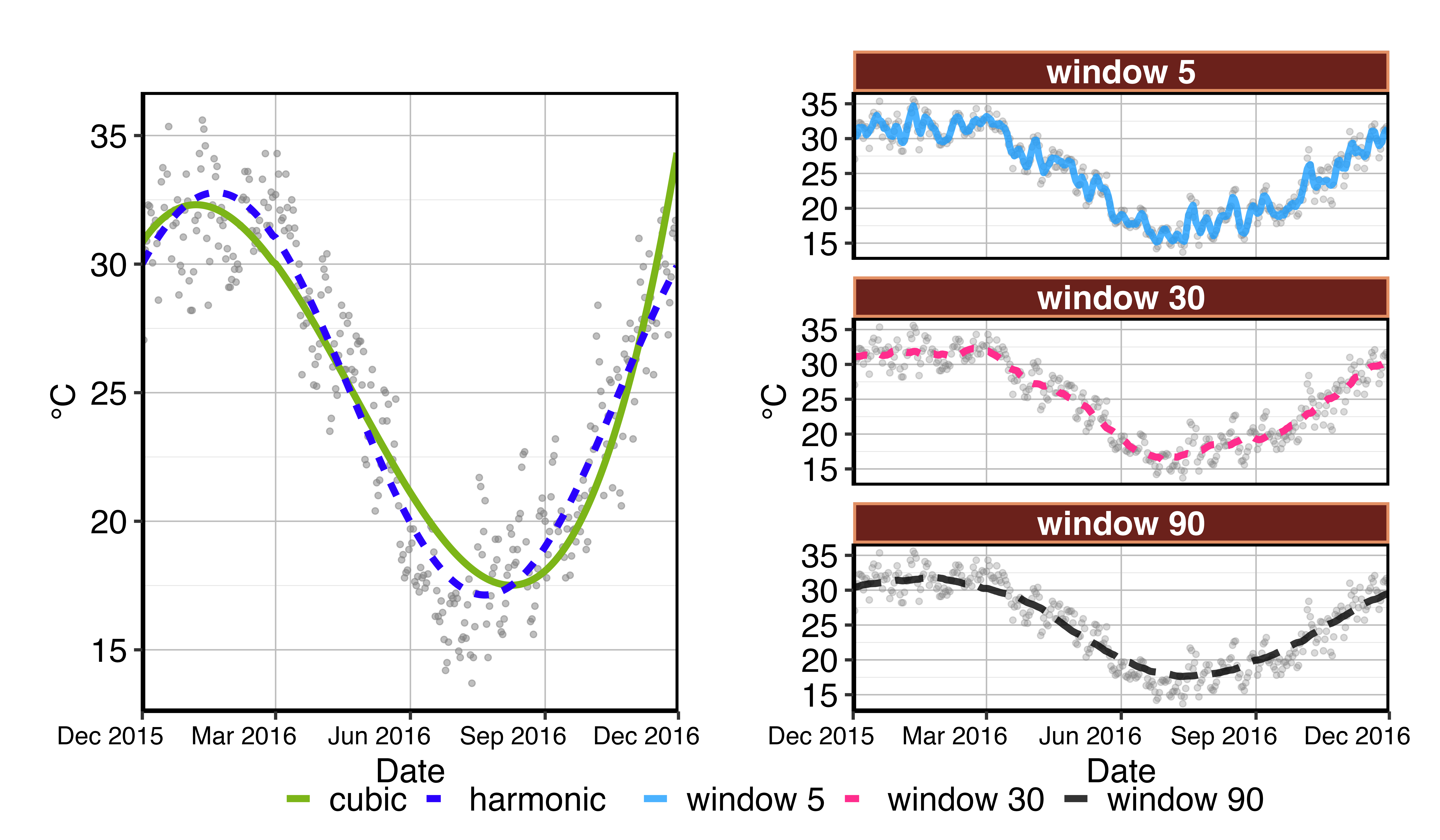}
 \caption{Common smoothing approaches applied to the 2016 daily median maximum temperatures (December 2015 to November 2016). Left: cubic polynomial and harmonic fits. Right: centred moving averages with 5, 30, and 90 day windows. Grey points show the daily median raw values.}
 \label{fig:Fig3}
\end{figure}

Fig.~\ref{fig:Fig3} displays the smoothing choices using a single example that we carry through the remainder of this section: the 2016 seasonal year of the Australia-wide daily median maximum temperature series (December 2015 to November 2016). For each date, there are 91 daily maximum temperature values, one from each location. We take the median across these 91 values to obtain one Australia-wide temperature value per day, producing a single annual curve that is then smoothed and analysed.

To mirror common practice, Fig.~\ref{fig:Fig3} compares smoothing approaches used in the seasonal literature. The left panel shows a cubic polynomial fit and a harmonic fit, reflecting approaches such as those used by \citet{alghamdi2024} and \citet{weller2020}. The right panel shows centred moving averages with 5, 30, and 90-day windows, reflecting common fixed-window choices in earlier work \citep{pena2015, kutta2016, hekmatzadeh2020, kotsias2021, yiqi2023, fan2025}. All of these approaches can produce a plausible annual curve for a single year, but they differ in how strongly they smooth the data and in the seasonal features they preserve.

The moving averages show the most direct trade-off between window length and performance. The 5-day window tracks short-lived fluctuations too closely and retains local peaks and troughs that are difficult to interpret as stable seasonal structure. The 30-day window produces a smoother annual pattern, but the timing and shape of the warm-season peak and cool-season trough are already altered. With a 90-day window, the series is clearly over-smoothed: the highest and lowest parts of the annual cycle are flattened, and both their values and timings are shifted away from the underlying seasonal pattern. The cubic polynomial imposes a rigid global shape and under-represents the depth of the winter trough. At the same time, the harmonic fit captures the broad cyclic pattern more plausibly but can still smooth away asymmetric shoulder behaviour and broad seasonal plateaus.

The key issue, therefore, is not simply whether one uses a moving average, a polynomial, or a harmonic form. Rather, it is whether the smoothing level is appropriate for the year and the seasonal structure under study. Two related matters follow:
\begin{itemize}
	\item different smoothing levels can yield different seasonal timings and estimated temperatures;
    \item a single fixed choice is unlikely to remain adequate across many years, and climates, especially as variability and seasonal shape evolve through time \citep{wu2025}.
\end{itemize}
What we require is not another single smoothing rule, but an approach that can examine the full range of smoothing levels and then determine which level is most appropriate for the seasonality problem at hand. The SiZer framework \citep{sizer1999} provides this perspective by examining over a range of smoothing levels, how evidence for features such as increasing, decreasing, and flat behaviour changes across the levels. With SALLI, we modify this idea by using the known structure of the annual temperature cycle to select a single optimal smoothing parameter for each year. The next section briefly summarises SiZer before showing how we alter it for seasonal-point estimation.

\subsection{SiZer}\label{sec:sizer}

We briefly summarise the SiZer method here, as it provides the statistical basis that we adapt for our seasonal problem; for full technical details, see \citet{sizer1999}. SiZer is based on a scale-space perspective: scale refers to the amount of smoothing applied, indexed by the bandwidth ($h$), and space refers to the location ($x$) along the observed series. Instead of asking whether one particular smoothed curve is adequate, SiZer asks which features of the data are supported by statistical evidence across a range of bandwidths. 

For each $h$, SiZer smooths the same raw data series and then, using statistical significance testing, examines whether the resulting smoothed curve is increasing, decreasing, or approximately flat at each location. Let \(\{(x_i,y_i):i=1,\dots,T\}\) denote a univariate series with ordered time index \(x_i\) and response \(y_i\). For each bandwidth \(h>0\), we estimate a locally smoothed curve by local linear regression with Gaussian kernel weights,
\begin{equation}
\hat f_h(x)
=
\arg\min_{a,b}
\sum_{i=1}^{T}
\big[y_i-\{a+b(x_i-x)\}\big]^2 K_h(x_i-x),
\end{equation}
where
\begin{equation}
K_h(u)=\frac{1}{h}K\!\left(\frac{u}{h}\right),
\qquad
K(u)=\frac{1}{\sqrt{2\pi}}\exp(-u^2/2).
\end{equation}
Here, \(\hat f_h(x)\) is the fitted intercept at location \(x\), and the local slope
\begin{equation}
\hat f_h'(x)=\hat b(x;h)
\end{equation}
provides the derivative estimate at that $x$-value and bandwidth. The important point is that \(h\) controls the amount of smoothing through the kernel weights: small \(h\) values retain short-scale fluctuations, whereas large \(h\) values emphasise broader-scale structure. SiZer explores from the outset many such values \(h\) rather than fixing a single smoothing level.

\begin{figure}[ht]
 \centering
 \includegraphics[width=1\textwidth]{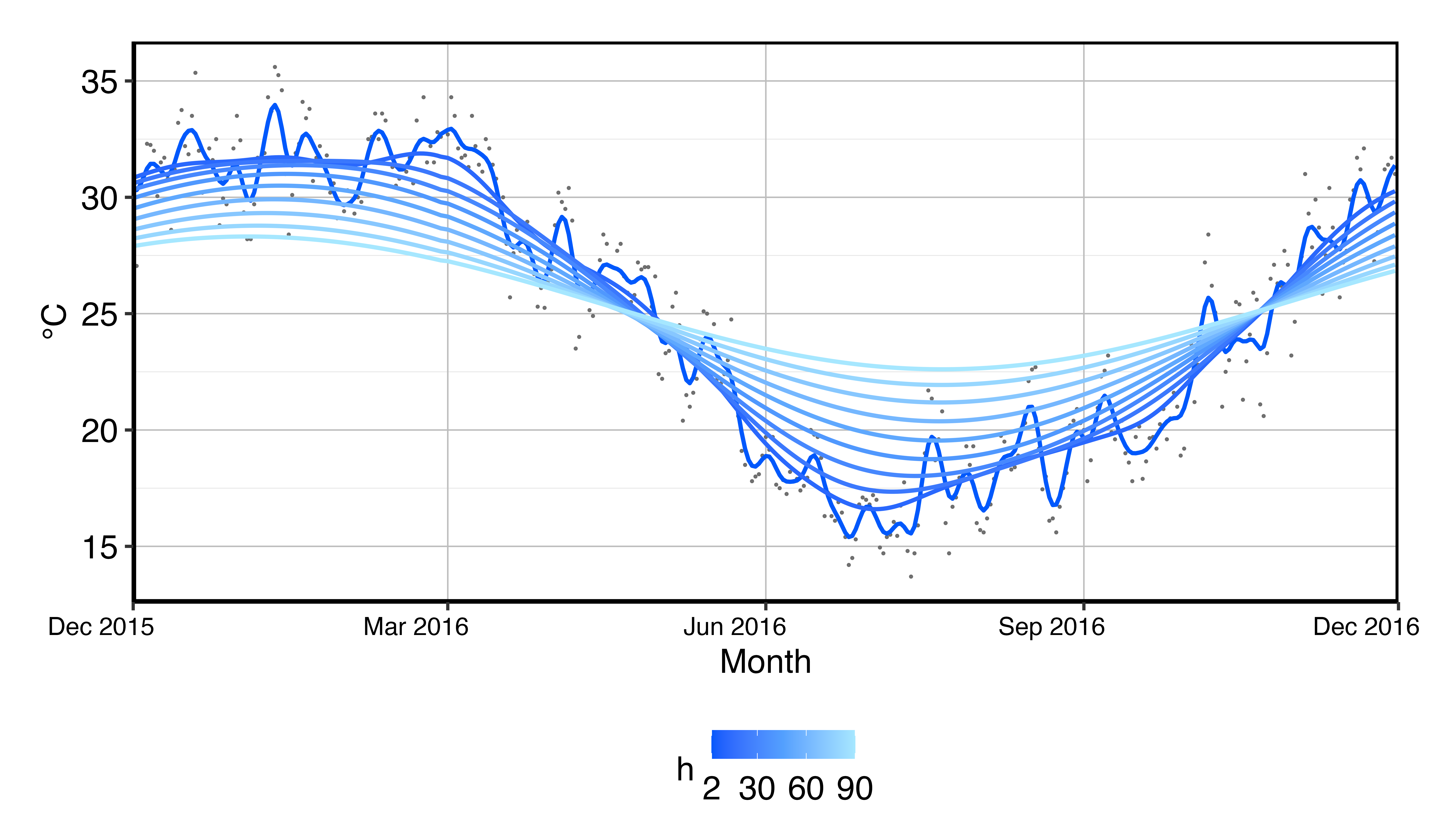}
 \caption{Fitted curves $\hat f_h(x)$ for the 2016 daily median maximum temperature series, using 15 bandwidths between 2 and 90 days. The colour legend shows $h$ in days on a linear scale. Darker blue curves correspond to smaller bandwidths, and lighter blue curves to larger bandwidths.}
 \label{fig:Fig4}
\end{figure}

Fig.~\ref{fig:Fig4} gives a visual overview of this idea for one temperature series. The figure shows fitted curves \(\hat f_h\) obtained over a range of bandwidths.

Reading Fig.~\ref{fig:Fig4} from small to large \(h\), we see how the interpretation of the series changes as more smoothing is applied. At small bandwidths, the fitted curves closely track the daily values and exhibit many local peaks and troughs. As the bandwidth increases, these short-scale variations are smoothed away. The fitted curves become progressively smoother, and the broader movements correspond more closely to the main decline from summer towards winter and the subsequent rise back towards the following summer. At still larger bandwidths, the fitted curves become flattened: peaks are pulled down, troughs are pushed up, and the timing of turning points may shift as the fit becomes dominated by coarse-scale structure. In the limit, if \(h\) is taken sufficiently large, the fitted curve approaches a straight line and the seasonal features of interest are no longer represented.

This SiZer view is useful because it lets us examine the full range of smoothing levels and the features that exist at each level rather than relying on a single fixed choice. A remaining question is how SiZer determines which parts of these smoothed curves represent statistically meaningful increases, decreases, or flat regions. This is done through significance testing of the derivative at each $x$-value and bandwidth.

SiZer makes inferences about the derivative of the smoothed temperature curve at each $(x,h)$-pair, rather than about the derivative of the underlying unsmoothed function. Using the significance level $\alpha=0.05$, SiZer constructs at each $(x,h)$-pair a confidence interval of the form
\begin{equation}
\label{eq:sizer-ci}
\hat f_h'(x)\ \pm\ q(h)\,\mathrm{se}\bigl(\hat f_h'(x)\bigr),
\end{equation}
where $q(h)$ is the critical value corresponding to $\alpha=0.05$ and $\mathrm{se}\bigl(\hat f_h'(x)\bigr)$ is the estimated standard error of the local slope estimate $\hat f_h'(x)$. This standard error captures the variability in the slope estimate, while the critical value controls the degree of confidence or uncertainty.

SiZer uses an adjusted critical value $q$ rather than $1.96$, the usual standard normal critical value for a two-sided $95\%$ pointwise confidence interval. For the joint inference at each $(x,h)$, the choice $1.96$ would increase the probability of identifying increases or decreases in the slope by chance. For this reason, SiZer uses a bootstrap approach to find a $q$-value which provides an approximate simultaneous adjustment for pairs $(x,h)$ and for multiple independent comparisons for each bandwidth $h$. This $q$ is typically larger than $1.96$, and is crucial in the SiZer confidence interval in Eq.~\eqref{eq:sizer-ci} or, equivalently, in SiZer hypothesis tests relating to significant increases, decreases and flat regions of the smoothed estimate $\hat f_h$.

The bootstrap approach and significance testing are integral parts of SiZer; for more details, we refer the interested reader to \citet{sizer1999}.

\begin{figure}[ht]
 \centering
 \includegraphics[width=1\textwidth]{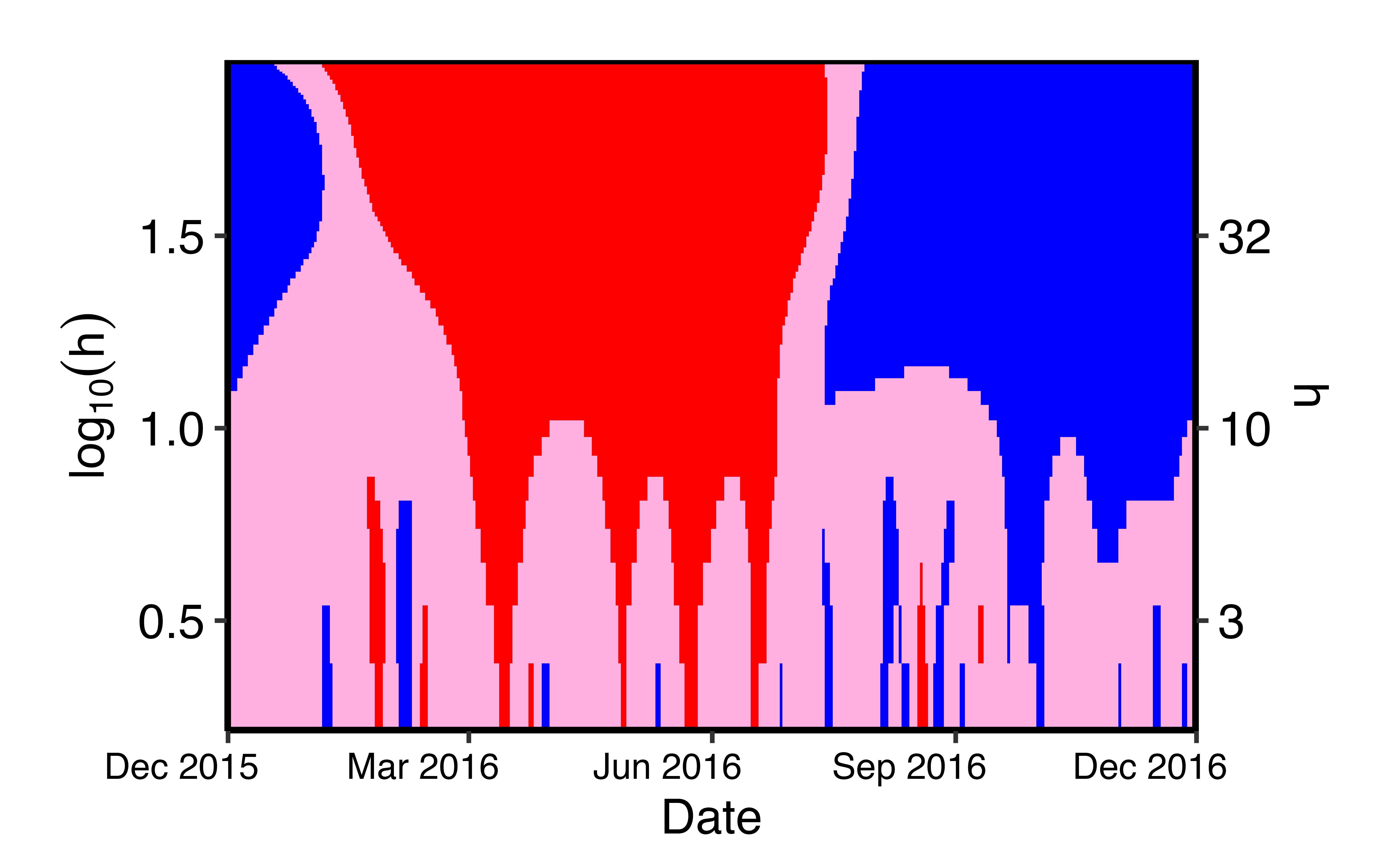}
 \caption{SiZer map for the 2016 daily median maximum temperature series. The left vertical axis is \(\log_{10}(h)\), and the right axis shows the corresponding bandwidth values \(h\).}
 \label{fig:Fig5}
\end{figure}

Fig.~\ref{fig:Fig5} visualises these tests through the SiZer map for the same series as shown in Fig.~\ref{fig:Fig4}. The map summarises the behaviour of the estimated derivative across two dimensions: location \(x\) and scale \(h\). Each pixel corresponds to a specific pair \((x,h)\) and is coloured according to the sign and significance of the derivative at that \((x,h)\). Blue if the interval lies above zero (significant increase in $\hat f_h$), red if the interval lies below zero (significant decrease in $\hat f_h$), and pink if the interval contains zero (no statistically significant increase or decrease). We display the bandwidth on the $y$-axis on a \(\log_{10}(h)\) scale on the left, while the right axis reports the corresponding bandwidth values \(h\).

Reading the map from bottom to top (small to large \(h\)), we see how the interpretation of the estimated $\hat f_h$ changes as the smoothing level increases. At small bandwidths (e.g. $h = 3$), the map is dominated by pink with short, scattered stretches of blue and red. This reflects the fact that, at very fine scales, day-to-day variability produces many short-lived ups and downs that do not stand out as clear increasing or decreasing trends. As the bandwidth increases (e.g. $h = 10$), these short-scale variations are smoothed away, and the map begins to show longer, more coherent runs of blue and red separated by pink transition bands. These pink regions are especially important because they mark locations where the derivative is not distinguishable from zero, that is, where the smoothed curve is locally flat at that bandwidth. Reading across a fixed bandwidth, left to right, a change from blue to pink to red indicates that the fitted curve rises, becomes locally flat, and then falls; this corresponds to a peak. Similarly, a change from red to pink to blue indicates that the fitted curve falls, becomes locally flat, and then rises; this corresponds to a trough.

Taken together, the fitted curves and the SiZer map provide evidence for increasing, decreasing, and flat behaviour across bandwidths. Rather than committing to a single smoothing choice upfront, we examine the full range of \(h\) values and see which aspects of the annual cycle remain supported as the amount of smoothing varies. For our problem, this is the key starting point. The next section shows how we exploit this idea to define a data-driven smoothing level appropriate for each yearly seasonal curve, and then use that choice to characterise summer, autumn, winter, and spring midpoints.

\subsection{Seasonal Adaptive LocalisabLe Index (SALLI)}\label{sec:sizerSeasonality}

We now describe how we use SiZer to determine the optimal bandwidth for estimating seasonal midpoints and their temperatures for each yearly cycle. Recall that the SiZer smooths local features that exist at a given bandwidth $h$. SALLI uses domain knowledge of the climate to specify the seasonal structure of interest. For the climates considered here, this comprises one broad summer peak and one winter trough per annual cycle. The required sign sequence can be adapted where domain knowledge indicates a different annual structure.

We denote the outcomes of the SiZer significance tests for each $h$ by an ordered sequence of signs, with blue, pink, and red corresponding to $+1$, $0$, and $-1$, respectively. After collapsing adjacent cells of the same sign into continuous runs, the required seasonal pattern is
\begin{equation}
\label{eq:order}
(+1,0,-1,0,+1,0,-1).
\end{equation}
The second $+1,0,-1$ transition is included because the extended window captures the following summer peak, needed to estimate the spring midpoint. The full sequence represents an increase into summer, a summer turning region, a decrease towards winter, a winter turning region, an increase into the following summer, its turning region, and the subsequent decrease. Pink regions are therefore required between runs of opposite signs. An additional pink region within an increasing or decreasing segment does not satisfy this pattern.

Among the bandwidths that recover this structure, the SALLI bandwidth is chosen to support three objectives:
\begin{enumerate}
	\item Prioritise accurate estimates of the summer maximum and winter minimum temperatures by selecting the minimum smoothing needed to recover the seasonal pattern, limiting the flattening of peaks and troughs.
    \item Estimate the summer and winter midpoints and their fitted temperatures.
    \item Estimate the autumn and spring midpoints and their fitted temperatures.
\end{enumerate}

We examine the bandwidth grid from largest to smallest and retain the smallest value satisfying Eq.~\eqref{eq:order}, denoted $h^\star$. This minimum is particularly important for temperature estimation: larger bandwidths may retain the required ordering while lowering the fitted summer maximum and raising the winter minimum. For the 2016 example considered here, the selected bandwidth is
\begin{equation}
\label{eq:hstar2016}
h^\star=15\ \text{days},
\end{equation}
as illustrated in Fig.~\ref{fig:Fig7}.

To show how this procedure is applied, consider a target year $Y$. We use observations from 1~August $(Y-1)$ to 31~May $(Y+1)$, giving $T=669$ days after excluding leap days. This adds five calendar months before and after the target calendar year, placing the summer and winter turning regions away from the window boundaries and including the following summer peak needed to identify spring. Exploratory checks of extensions from one to twelve months supported this choice as a balance between resolving the initial rise towards summer and limiting the influence of neighbouring cycles.

\begin{figure}[ht]
 \centering
 \includegraphics[width=1\textwidth]{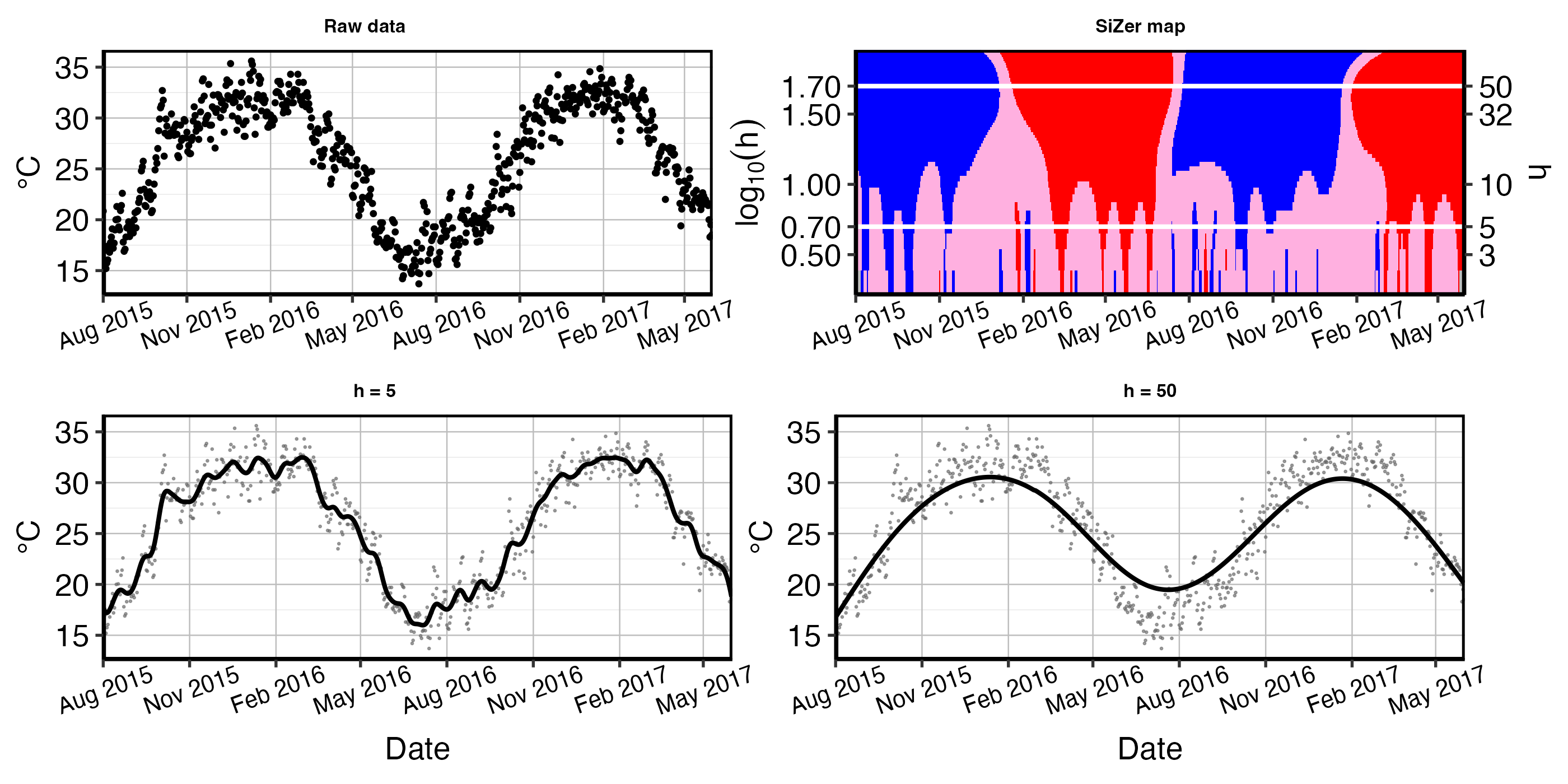}
 \caption{Extended-window analysis for the 2016 seasonal year using the Australia-wide daily median maximum temperature series. Top left: raw daily median temperatures from 1~August~2015 to 31~May~2017. Top right: corresponding SiZer map. Horizontal white lines mark two illustrative bandwidths, $h=5$ and $h=50$. Bottom left: fitted curve $\hat f_h(x)$ at $h=5$, overlaid on the raw data. Bottom right: fitted curve $\hat f_h(x)$ at $h=50$, overlaid on the raw data.}
 \label{fig:Fig6}
\end{figure}

\begin{figure}[ht]
 \centering
 \includegraphics[width=1\textwidth]{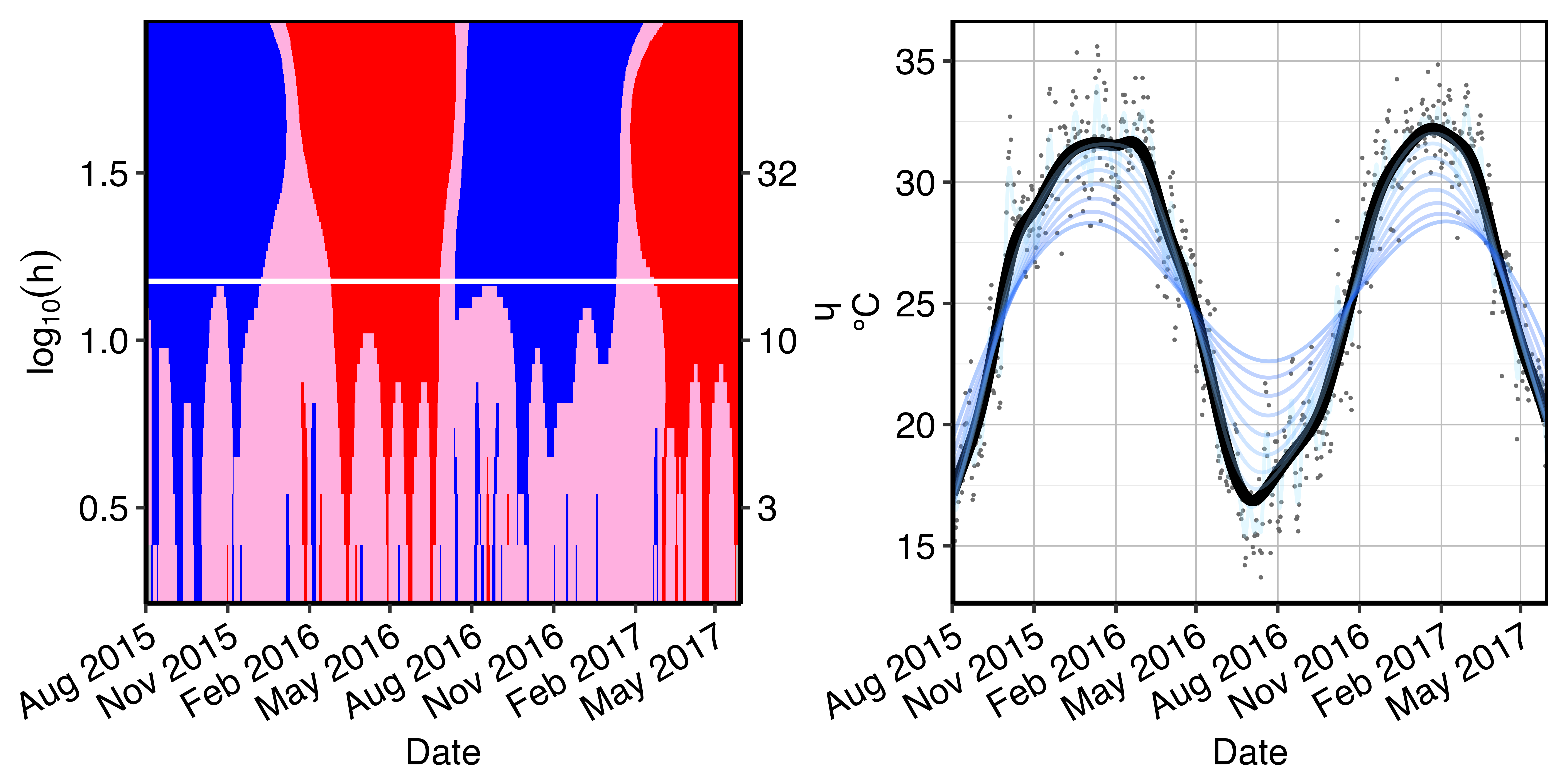}
 \caption{(Left) SiZer map for the 2016 daily median maximum temperature series. The white horizontal line marks the data-driven bandwidth choice, $h^\star=15$. (Right) Fitted curves $\hat f_h(x)$ over a range of bandwidths. The black curve shows $\hat f_{h^\star}$, overlaid on the raw daily median maximum temperatures (points).}
 \label{fig:Fig7}
\end{figure}

Fig.~\ref{fig:Fig6} illustrates the extended-window analysis using the Australia-wide daily median maximum temperature series. The upper panels show the raw series and corresponding SiZer map, while the lower panels compare fitted curves at $h=5$ and $h=50$. At $h=5$, the fitted curve retains multiple local peaks and troughs, and the required seasonal pattern in Eq.~\eqref{eq:order} is not recovered. At $h=50$, the required pattern is present, but the SiZer map shows that smaller bandwidths, including $h=49, 48, \ldots $, also recover this structure. The fitted curve at $h=50$ therefore illustrates that recovering the seasonal ordering alone is insufficient for selecting the bandwidth. Its summer peaks are lowered, and its winter trough is raised by smoothing, affecting the temperature estimates. Moving down through the acceptable bandwidths allows us to retain the required structure while flattening these features less. This motivates selecting the minimum bandwidth, $h^\star=15$ in Eq.~\eqref{eq:hstar2016}, with particular attention to preserving the fitted summer and winter temperature levels.

Fig.~\ref{fig:Fig7} illustrates why the selection stops at $h^\star=15$, marked by the white horizontal line in the left panel. At the tested bandwidth immediately below this value, the second blue run is interrupted by an additional pink patch. The resulting sign sequence therefore contains an additional flat region within the rise towards the following summer and does not match Eq.~\eqref{eq:order}. At $h^\star=15$, this internal pink patch disappears, leaving a continuous blue run. The right panel shows fitted curves $\hat f_h$ over a range of bandwidths, with $\hat f_{h^\star}$ highlighted in black. This comparison shows how the selected curve removes short-scale fluctuations while retaining the broad seasonal features, and how further smoothing flattens the peaks and troughs. The bandwidth choice is determined by applying the predefined seasonal-pattern criterion to the results of the SiZer significance tests and is reproducible for the same data, bandwidth grid and inferential settings.

We next turn to item two on the list of properties of SALLI to establish the summer and winter midpoints. A naive approach to defining the summer and winter midpoints would be to take the global maximum and minimum of $\hat f_{h^\star}$. However, for temperature data, there is rarely a single day that can be identified as the year's maximum or minimum. Warm and cool conditions typically persist, so the peak and trough are viewed as periods in which the fitted curve is approximately flat near its highest or lowest levels. Recalling that the SiZer \(h^\star\) includes a hypothesis for each $(x,h)$, we use the fact that the pink band indicates the flat interval, where no significant increase or decrease occurs. 

For the optimal \(h^\star\) we therefore define the summer midpoint as the midpoint in time of the pink interval, lying between an increasing run and the following decreasing run. The winter midpoint is the midpoint in time of the pink interval, lying between a decreasing run and the following increasing run.

\begin{figure}[ht]
  \centering
  \includegraphics[width=1\textwidth]{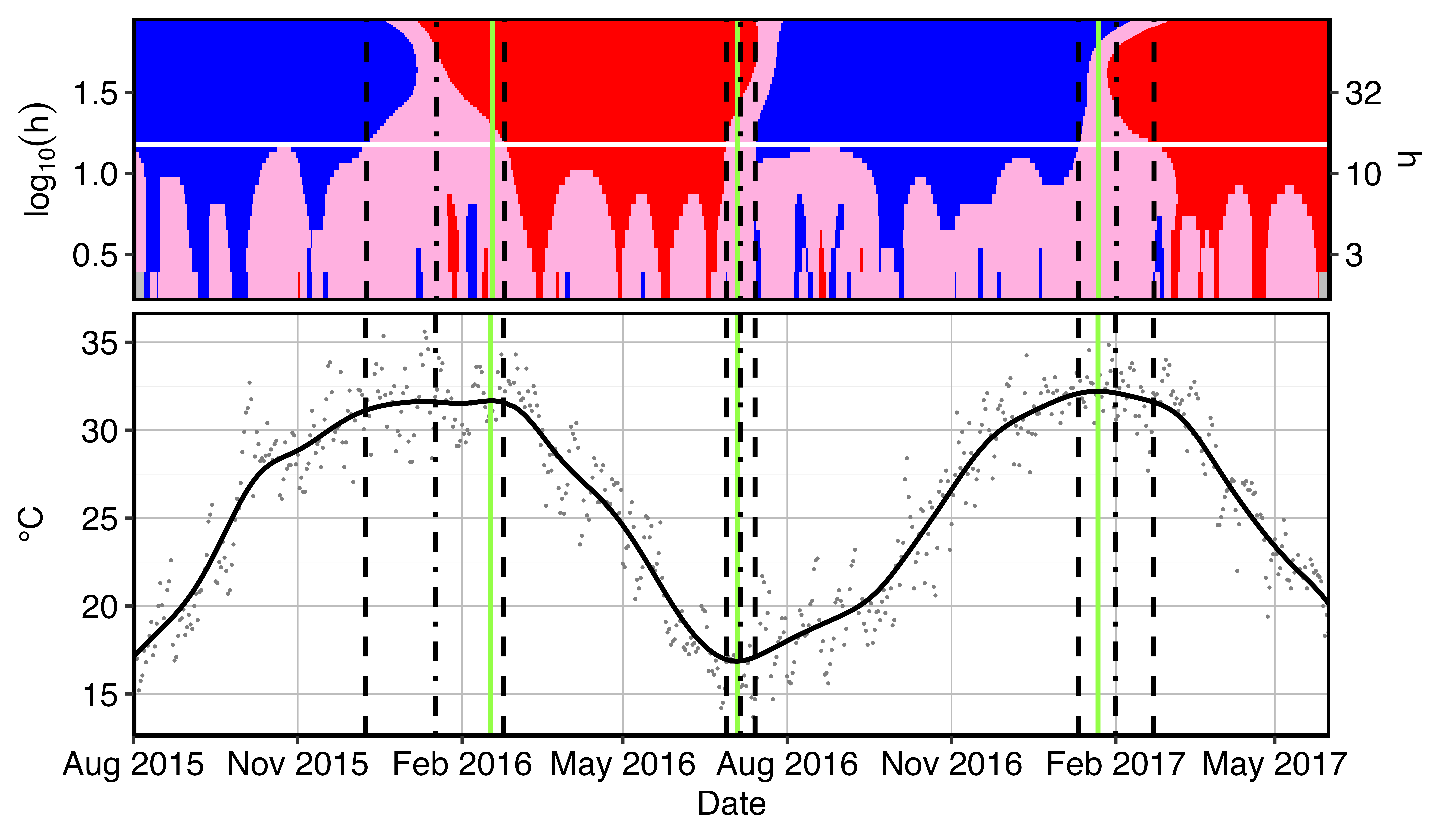}
\caption{(Top) SiZer map with $h^\star=15$ marked by the grey horizontal line. (Bottom) Fitted curve overlaid on the raw data. In both panels, light green vertical lines mark the fitted summer maxima and winter minimum, black dashed lines mark the start and end of each pink SiZer transition interval, and black dash-dotted lines mark the midpoints in time of these intervals, used as the summer and winter midpoint estimates.}
  \label{fig:Fig8}
\end{figure}

Fig.~\ref{fig:Fig8} illustrates this at $h^\star=15$. Light green vertical lines mark the locations of the peaks and trough of $\hat f_{h^\star}$, black dashed lines mark the bounds of each pink flat interval, and dashed-dotted black lines indicate our estimate of the summer and winter midpoints. The lower panel shows the same vertical markers superimposed on the smoothed curve and the raw data. Around the summer peak, the Australia-wide 2016 series exhibits sustained high temperatures from early December 2015 through to early March 2016. Over this interval, many daily values exceed $30^\circ$C, and the fitted curve remains essentially flat, based on significance testing at $h^\star$, indicating a broad warm-season plateau rather than a single dominant day. This is reflected in the SiZer map by a wide pink band between the blue and red regions. If we were to take the naive approach of using the single highest-fitted day (light green line), the estimate would fall near the end of February. By contrast, the midpoint of the pink interval, 17 January (dashed-dotted black line), provides a representative centre of the sustained warm period, matching our aim to summarise the timing of the warm-season period rather than an arbitrary single-day extreme. Turning to the winter trough, the coldest period is much more concentrated. The pink interval between the red and blue regions is narrow, and its midpoint lies in early July. In this case, the midpoint closely matches the day of the minimum of $\hat f_{h^\star}$. When the trough (or peak) is sharp, our midpoint-based definition agrees with the conventional global max and min. When the feature is wide, it yields a more stable time estimate.

We now turn to two inflection points of autumn and spring, the third item on the list of properties of SALLI. In principle, inflections could be estimated using a second-derivative SiZer \citep{chaudhuri2002}. In our context, however, the second derivative is very sensitive to small-scale variations, requiring larger bandwidths and leading to over-smoothed features. This behaviour is noted by \citet{chaudhuri2002}, who provide a mathematical explanation for why the variance of the second-derivative estimate increases rapidly at small scales, thereby inflating confidence intervals and masking significance. We therefore define autumn and spring using data-driven temperature-based transition points.

Having identified the maxima and minima, we define the autumn and spring midpoints as the time points where the fitted curve reaches the midpoint temperature between the summer maximum and winter minimum. The autumn midpoint is taken from the descent between the first summer midpoint and the winter midpoint, while the spring midpoint is taken from the ascent between the winter midpoint and the second summer midpoint in the following year. This construction uses the smooth monotone segments identified by \(\hat f_{h^\star}\), avoids taking higher derivatives of a noisy curve, and links the transition times directly to the heights of the summer and winter turning points.  

\begin{figure}[ht]
 \centering
 \includegraphics[width=1\textwidth]{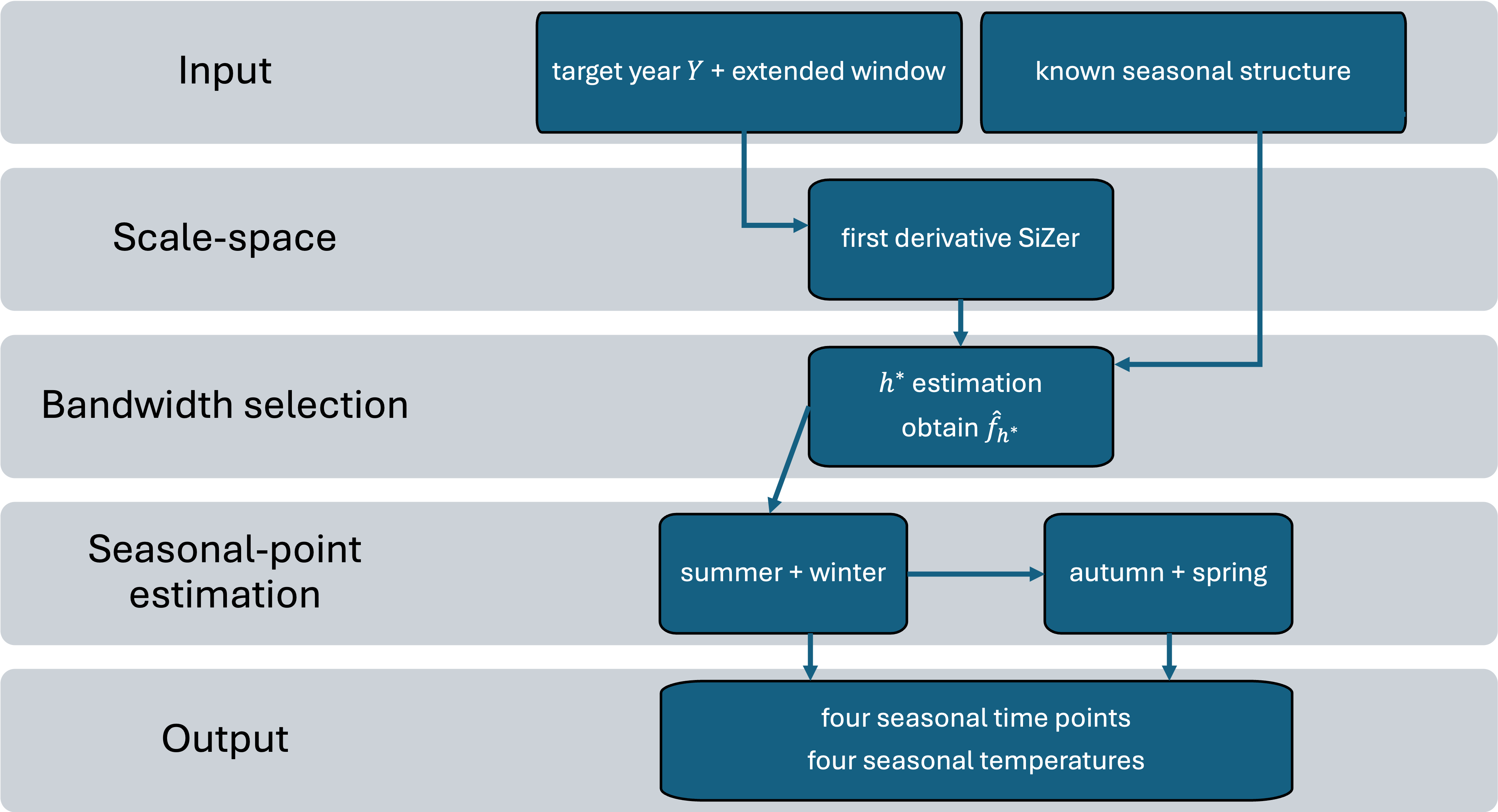}
 \caption{Overview of the SALLI workflow.}
 \label{fig:Fig9}
\end{figure}

Fig.~\ref{fig:Fig9} summarises the full SALLI workflow. The procedure combines the extended-window construction, first-derivative SiZer, data-driven bandwidth selection, and estimation of seasonal timing and temperatures. The output is a single optimal bandwidth \(h^\star\), the corresponding fitted curve \(\hat f_{h^\star}\), and eight annual index values: four seasonal timings and the fitted temperatures at those four points.

For a target seasonal year \(Y\) and a temperature series \(\{(t_i,y_i)\}_{i=1}^{T}\), the SALLI procedure is:

\begin{enumerate}
\item Construct the extended window using observations from 1~August \((Y-1)\) to 31~May \((Y+1)\). This avoids edge effects when the warm-season peak falls at the end of the calendar year and ensures that the following summer peak is available for estimating the spring point.

\item Compute the first-derivative SiZer map for the extended series over a grid of bandwidths \(h \in [2,90]\).

\item For each bandwidth, reduce the SiZer row to its ordered sign structure by collapsing adjacent cells of the same sign into continuous runs, with increasing, flat, and decreasing regions represented by \(+1\), \(0\), and \(-1\), respectively.


\item Select $h^\star$ as the smallest bandwidth that recovers the required summer--winter--summer pattern $(+1,0,-1,0,+1,0,-1)$. This limits additional smoothing that can lower fitted summer peaks and raise the winter trough.

\item Retain the corresponding fitted curve \(\hat f_{h^\star}\).

\item Define the summer and winter midpoints from the SiZer plateau regions at \(h^\star\). The midpoint of the pink band between a \(+1\) run and a \(-1\) run defines the summer point, and the midpoint of the pink band between a \(-1\) run and a \(+1\) run defines the winter point. The second \(+1 \rightarrow -1\) transition is retained only to support estimation of the spring point.

\item Define the autumn and spring midpoints from \(\hat f_{h^\star}\). These are the times on the decreasing and increasing segments, respectively, at which the fitted curve reaches the midpoint temperature between the summer and winter fitted levels.

\item Return the four annual seasonal-point estimates,
\[
\text{summer},\ \text{autumn},\ \text{winter},\ \text{spring},
\]
together with the fitted temperatures at those points.
\end{enumerate}

\begin{figure}[ht]
 \centering
 \includegraphics[width=1\textwidth]{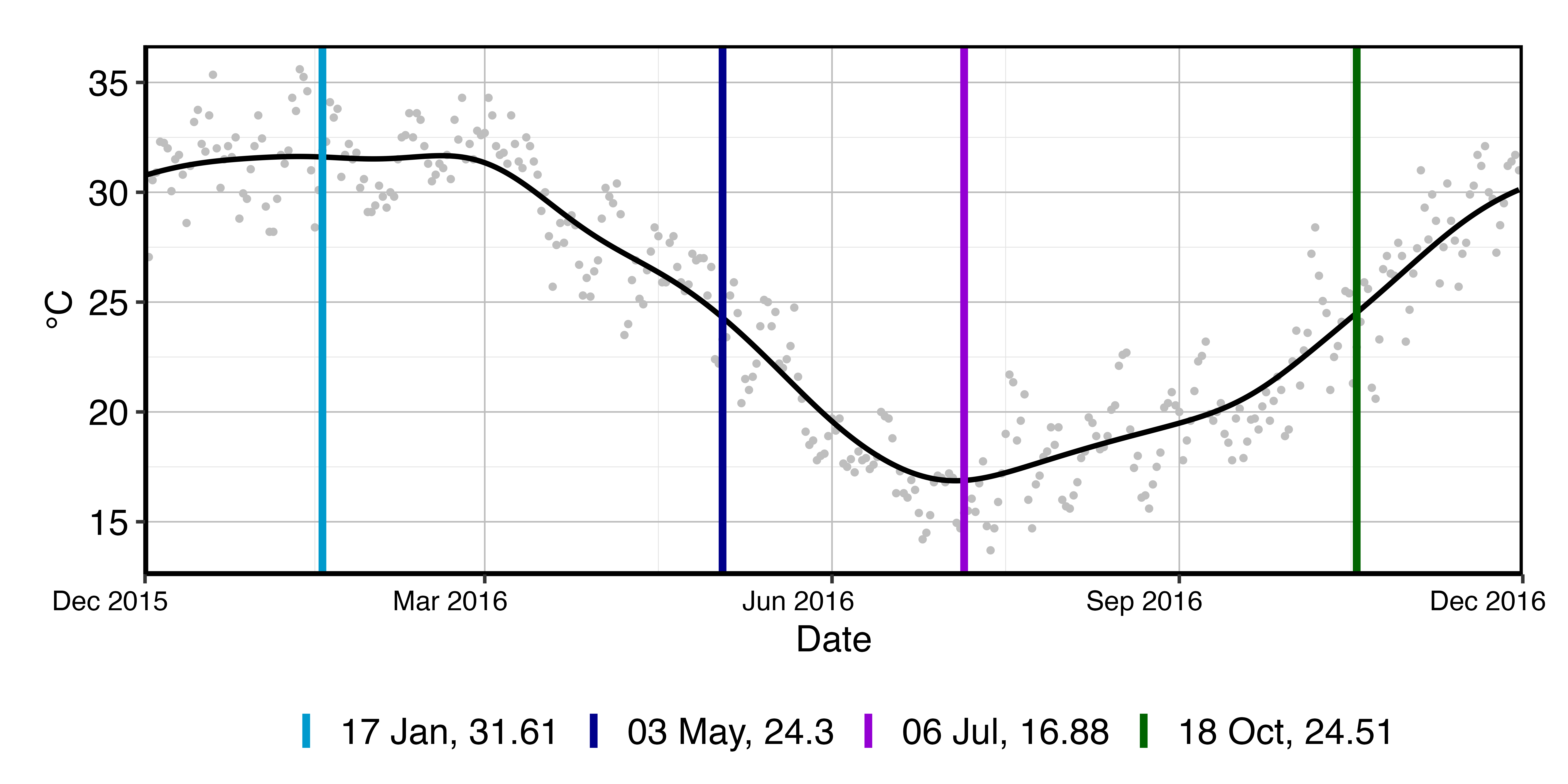}
 \caption{Grey points: daily values; black line: fitted \(\hat f_{h^\star}\), where \(h^\star=15\). Vertical markers show the estimated summer midpoint (light blue), autumn midpoint (dark blue), winter midpoint (purple), and spring midpoint (green). Legend entries give the estimated midpoint dates and fitted temperatures ($^\circ$C) for the 2016 seasonal year.}
 \label{fig:Fig10}
\end{figure}

Applied to the 2016 daily median maximum series, this procedure yields the four seasonal midpoints shown in Fig.~\ref{fig:Fig10}. For this example, the estimated summer midpoint is 17~January at \(31.61^\circ\)C, the autumn midpoint is 3~May at \(24.30^\circ\)C, the winter midpoint is 6~July at \(16.88^\circ\)C, and the spring midpoint is 18~October at \(24.51^\circ\)C. Fig.~\ref{fig:Fig10} shows the end product of the SALLI workflow: four unambiguous points on the annual temperature curve, each with both a time estimate and an associated temperature estimate.

\section{Results}
\label{sec:results}

This section illustrates how SALLI characterises changes in Australia's annual maximum-temperature cycle from 1966 to 2023. We focus on the Australia-wide series to demonstrate the index's performance. The analyses in this section are based on daily medians of maximum temperature across the 91 locations. For notational convenience, we often omit the word ``median'' and refer simply to national or Australia-wide temperatures. The comparisons below are summaries of seasonal timing and temperature across the selected periods.

\begin{figure}[ht]
 \centering
 \includegraphics[width=1\textwidth]{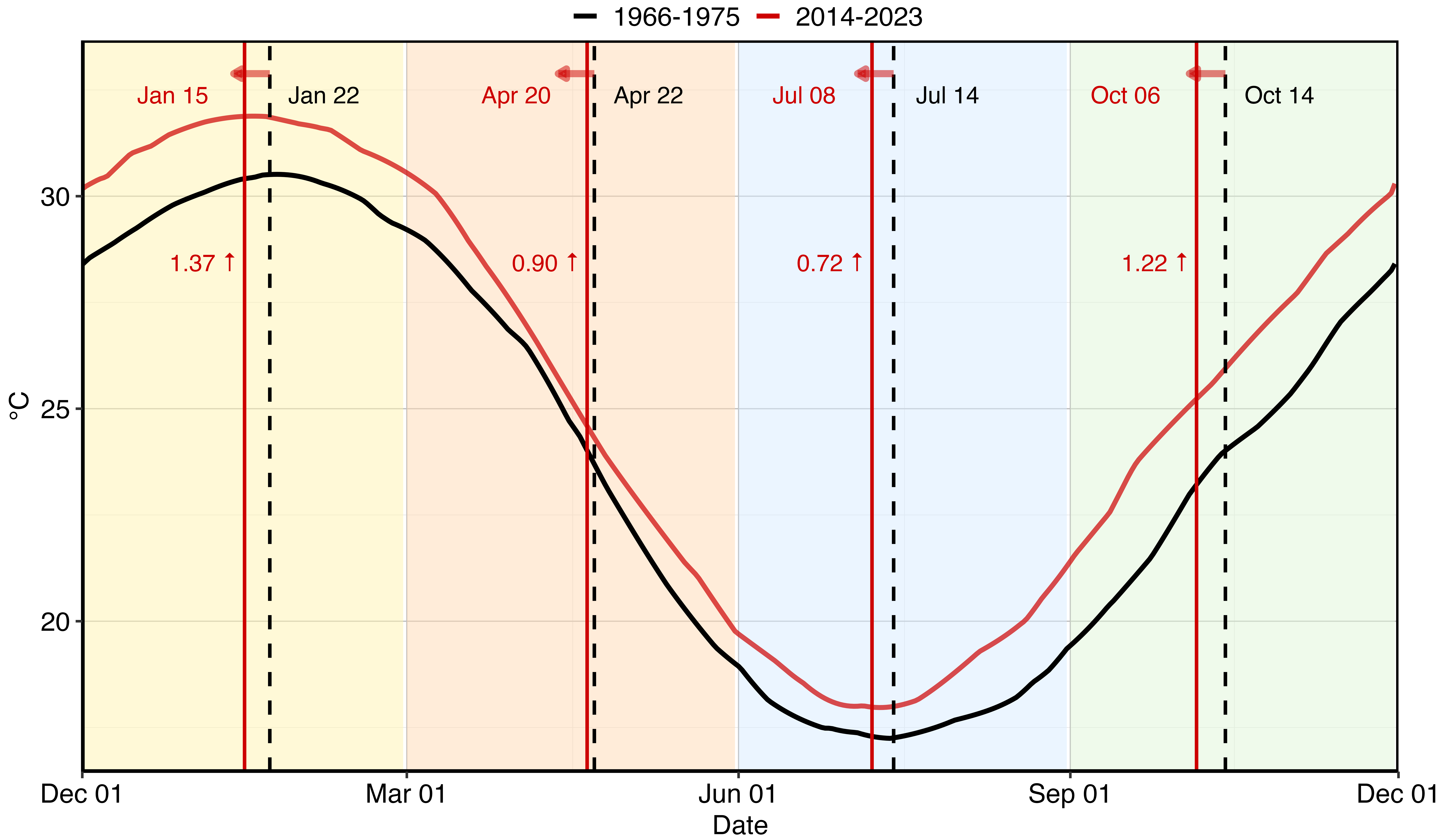}
\caption{Australia-wide seasonal-cycle shifts between the first decade (1966--1975; black) and the last decade (2014--2023; red), for medians of daily maximum temperature from 91 stations. Curves are decade-median SALLI-smoothed annual cycles, where each year is smoothed at its own selected bandwidth $h^\star$. Vertical lines mark the decade-average timing of the four seasonal midpoints. Temperature changes are calculated by evaluating each curve at its own decade-average midpoint date and subtracting the earlier decade's fitted temperature from the later decade's value.}
 \label{fig:Fig11}
\end{figure}

Fig.~\ref{fig:Fig11} summarises Australia-wide changes in seasonal timing for daily maximum temperature. For each year from 1966 to 2023, we construct one Australia-wide daily median maximum temperature series. To avoid edge effects when the warm season spans the end of the calendar year, and to allow estimation of the spring midpoint from the subsequent rise toward the next summer, each seasonal year is analysed within the extended window described in Sect.~\ref{sec:sizerSeasonality}. This is why the analysis begins in 1966 rather than 1965, and ends in 2023 rather than 2024. We apply SALLI to each annual Australia-wide median series to estimate four seasonal midpoints: summer, autumn, winter, and spring, together with their four temperatures.

In Fig.~\ref{fig:Fig11}, the black curve is formed from the 10 yearly SALLI-smoothed curves for 1966-1975. Each yearly curve is first estimated using its own selected bandwidth \(h^\star\). Then, for each day of the seasonal year, we take the median of the 10 fitted values to obtain the decade-median curve. The red curve is constructed in the same way for 2014-2023. The vertical lines mark the decade-average estimates for the four seasonal midpoints. This figure therefore compares decade-level summaries derived from annual SALLI estimates. For each seasonal point, we evaluate the difference between the two decade averages, similarly for the temperatures. This compares temperatures associated with the same seasonal feature, while allowing its timing to differ across decades.

The main Australia-wide results can be summarised as follows:
\begin{itemize}
    \item the seasonal cycle shifts to earlier in the year;
    \item temperatures at the seasonal midpoints increase;
    \item there is a shortening of the cooler half of the annual cycle;
\end{itemize}

The dominant Australia-wide result is a shift to earlier times of the seasonal midpoints. Relative to the first decade, the summer, winter, and spring midpoints all occur earlier in the most recent decade, indicating a broad phase advance in the annual maximum-temperature cycle. The clearest timing change occurs for spring, which shifts about 8 days between the first and last decades. Summer and winter also advance by around 7 and 6 days, respectively, while autumn shows only a small shift of around 2 days.

These timing shifts coincide with temperature increases at all four seasonal midpoints. Summer shows the largest increase in maximum temperature, about $1.37~^\circ\mathrm{C}$, followed by spring at $1.22~^\circ\mathrm{C}$. Autumn and winter increase by about $0.90~^\circ\mathrm{C}$ and $0.72~^\circ\mathrm{C}$, respectively. The largest temperature increase and timing advance occur at different seasonal points: summer warms most, while spring moves most. Autumn shows comparatively little change in timing despite warming at its midpoint, illustrating that temperature and timing can change by different amounts.

Unequal shifts in the seasonal points imply changes in the lengths of seasonal segments, the duration of the cooler half of the annual cycle, defined here as the interval from autumn to spring. Because the spring midpoint advances more than the autumn midpoint (roughly 8 days versus 2 days), a cool season shortens by about 6 days in the first-versus-last decade comparison. Other segment lengths also change, indicating that the annual cycle shifts earlier overall and redistributes time across its cooling and warming phases.

The first-versus-last-decade comparison provides an overall summary, Table~\ref{tab:Tab1} shows how the estimates vary across consecutive five-year periods. Each period is summarised from the annual SALLI results, and changes between consecutive periods are accumulated from the initial period, 1966--1970. The table ends at 2016--2020 to keep all periods equal in length. The remaining three years, 2021--2023, are excluded from this table but retained in the decade comparison in Fig.~\ref{fig:Fig11}.

\begin{table}[!tp]
\centering
\scriptsize
\setlength{\tabcolsep}{4.5pt}
\renewcommand{\arraystretch}{1.04}
\caption{Cumulative changes in seasonal midpoint timing and temperature for maximum temperature across consecutive five-year periods. Each row gives the accumulated change through the period in the second column, relative to 1966--1970. Negative timing values indicate earlier midpoints, and positive temperature values indicate warming. Colours indicate the direction of change.}
\label{tab:Tab1}

\begin{tabular*}{\textwidth}{@{\extracolsep{\fill}} l l r r @{}}
\toprule
\textbf{Period start} & \textbf{Period end} &
\textbf{Days} &
\textbf{$^\circ$C} \\
\midrule

\multicolumn{4}{c}{\textbf{Summer}}\\
\midrule
1966--1970 & 1971--1975 & \deltasignedday{-6}  & \deltasignedtemp{-0.30} \\
1971--1975 & 1976--1980 & \deltasignedday{-7}  & \deltasignedtemp{0.64} \\
1976--1980 & 1981--1985 & \deltasignedday{-7}  & \deltasignedtemp{0.79} \\
1981--1985 & 1986--1990 & \deltasignedday{4}   & \deltasignedtemp{0.07} \\
1986--1990 & 1991--1995 & \deltasignedday{-10} & \deltasignedtemp{0.41} \\
1991--1995 & 1996--2000 & \deltasignedday{1}   & \deltasignedtemp{0.24} \\
1996--2000 & 2001--2005 & \deltasignedday{-7}  & \deltasignedtemp{0.67} \\
2001--2005 & 2006--2010 & \deltasignedday{-3}  & \deltasignedtemp{0.82} \\
2006--2010 & 2011--2015 & \deltasignedday{-9}  & \deltasignedtemp{0.59} \\
2011--2015 & 2016--2020 & \deltasignedday{-11} & \deltasignedtemp{1.68} \\

\addlinespace[2pt]
\midrule
\multicolumn{4}{c}{\textbf{Autumn}}\\
\midrule
1966--1970 & 1971--1975 & \deltasignedday{-1} & \deltasignedtemp{0.55} \\
1971--1975 & 1976--1980 & \deltasignedday{1}  & \deltasignedtemp{0.15} \\
1976--1980 & 1981--1985 & \deltasignedday{0}  & \deltasignedtemp{0.56} \\
1981--1985 & 1986--1990 & \deltasignedday{3}  & \deltasignedtemp{0.17} \\
1986--1990 & 1991--1995 & \deltasignedday{0}  & \deltasignedtemp{0.94} \\
1991--1995 & 1996--2000 & \deltasignedday{-2} & \deltasignedtemp{0.47} \\
1996--2000 & 2001--2005 & \deltasignedday{4}  & \deltasignedtemp{0.56} \\
2001--2005 & 2006--2010 & \deltasignedday{0}  & \deltasignedtemp{0.79} \\
2006--2010 & 2011--2015 & \deltasignedday{-3} & \deltasignedtemp{1.03} \\
2011--2015 & 2016--2020 & \deltasignedday{-2} & \deltasignedtemp{1.95} \\

\addlinespace[2pt]
\midrule
\multicolumn{4}{c}{\textbf{Winter}}\\
\midrule
1966--1970 & 1971--1975 & \deltasignedday{1}  & \deltasignedtemp{0.19} \\
1971--1975 & 1976--1980 & \deltasignedday{-3} & \deltasignedtemp{0.67} \\
1976--1980 & 1981--1985 & \deltasignedday{-5} & \deltasignedtemp{0.11} \\
1981--1985 & 1986--1990 & \deltasignedday{0}  & \deltasignedtemp{-0.44} \\
1986--1990 & 1991--1995 & \deltasignedday{-2} & \deltasignedtemp{0.63} \\
1991--1995 & 1996--2000 & \deltasignedday{-3} & \deltasignedtemp{0.36} \\
1996--2000 & 2001--2005 & \deltasignedday{-2} & \deltasignedtemp{0.90} \\
2001--2005 & 2006--2010 & \deltasignedday{-5} & \deltasignedtemp{0.66} \\
2006--2010 & 2011--2015 & \deltasignedday{-8} & \deltasignedtemp{0.62} \\
2011--2015 & 2016--2020 & \deltasignedday{-5} & \deltasignedtemp{1.38} \\

\addlinespace[2pt]
\midrule
\multicolumn{4}{c}{\textbf{Spring}}\\
\midrule
1966--1970 & 1971--1975 & \deltasignedday{3}   & \deltasignedtemp{0.07} \\
1971--1975 & 1976--1980 & \deltasignedday{1}   & \deltasignedtemp{0.08} \\
1976--1980 & 1981--1985 & \deltasignedday{0}   & \deltasignedtemp{0.25} \\
1981--1985 & 1986--1990 & \deltasignedday{-5}  & \deltasignedtemp{-0.20} \\
1986--1990 & 1991--1995 & \deltasignedday{2}   & \deltasignedtemp{0.36} \\
1991--1995 & 1996--2000 & \deltasignedday{-2}  & \deltasignedtemp{0.46} \\
1996--2000 & 2001--2005 & \deltasignedday{0}   & \deltasignedtemp{0.60} \\
2001--2005 & 2006--2010 & \deltasignedday{-9}  & \deltasignedtemp{0.77} \\
2006--2010 & 2011--2015 & \deltasignedday{-13} & \deltasignedtemp{1.29} \\
2011--2015 & 2016--2020 & \deltasignedday{-3}  & \deltasignedtemp{1.33} \\

\bottomrule
\end{tabular*}
\end{table}

The five-year summaries show that earlier seasonal timing becomes more persistent in the later periods. Winter midpoints occur earlier than in the 1966--1970 baseline from 1991--1995 onwards, with cumulative advances ranging from 2 to 8 days. For summer, all periods from 2001--2005 onwards occur earlier than the baseline, with cumulative advances of 7, 3, 9 and 11 days. Spring shows a similar pattern from 2006--2010 onwards, with cumulative advances of 9, 13 and 3 days, and some instabilities in the later decades. These summaries suggest that warming is already evident in earlier decades, while earlier seasonal midpoints become apparent mainly in the 2000s. 

Overall, the Australia-wide maximum-temperature analysis highlights three main findings. First, the annual cycle shifts to earlier starts, with the strongest movement in spring. Second, the temperatures associated with the seasonal midpoints increase, especially in spring and summer. Third, the unequal timing shifts alter the lengths of seasonal segments, including a shortening of the cooler half of the cycle. From a methodological perspective, these results show that SALLI captures both horizontal changes in timing and vertical changes in temperature in a single multivariate index. The five-year summaries show that the clearest timing changes emerge mainly in the 2000s and later.

\section{Discussion}
\label{sec:discussion}

We propose SALLI, a multivariate index for characterising thermal seasonality from daily temperature curves. SALLI uses the scale-space SiZer approach to determine an optimal smoothing bandwidth for each year that accurately estimates mid-seasonal timepoints and their temperatures. SALLI identifies four unique seasonal midpoints directly from the annual temperature cycle rather than defining seasons through fixed thresholds, percentiles, or calendar boundaries. By focusing on the annual cycle, seasonal change can be described in terms of the curve itself: the timing of the warm-season peak, the cool-season trough, and the two transition points between them. In this way, the process moves beyond asking whether temperatures rise, and instead asks how the annual seasonal cycle changes in timing, shape, duration and temperature.

SALLI also contributes methodologically through its data-driven smoothing procedure. Much of the existing literature adopts a single smoothing parameter or compares only a small number of smoothing choices, and then applies a single smoothing choice uniformly across all years or regions. Our approach uses SiZer and its derived statistical significance based on hypothesis tests to examine the full range of smoothing levels for each year, then defines a criterion that selects an optimal bandwidth \(h^\star\) appropriate for the seasonal structure of each yearly series. This is one of the main ways in which SALLI differs from other seasonal methods. It does not assume that a single smoothing parameter will remain adequate across decades of data. It also does not rely on arbitrary local thresholds whose interpretation can vary across climates. Instead, SALLI uses data-driven robust methods for choosing the optimal \(h^\star\) for each year. The resulting seasonal points are derived in a data-driven way, are directly tied to the observed annual curve and are statistically robust in terms of the smoothing scale that supports them.

The Australia-wide application shows that SALLI works well not only because it gives results consistent with existing seasonal-change literature, but because it identifies where on the annual temperature curve those changes occur. Existing studies in China, Korea, Europe, the Northern Hemisphere, the Southern Hemisphere, and Australia similarly report longer summers, shorter winters, earlier warm-season development, and relatively weaker or more variable changes in autumn and spring \citep{dong2010, pena2015, allen2016, kitowski2019, weller2020, wang2021, lee2022, yiqi2023, fan2025}. In this broad sense, our results are consistent with the established literature. In our companion paper, we show that SALLI applies and works equally well across different countries and hemispheres, as well as on the daily minimum temperature cycles. Threshold methods estimate the onset and duration of thermal conditions, while SALLI estimates seasonal midpoint timings and temperatures using consistent feature definitions and adaptive smoothing. These approaches provide complementary descriptions of seasonality, so differences between their dates alone do not establish that either performs better.

The results also show that seasonal change is not simply a uniform leftward translation of the annual temperature cycle. In the Australia-wide maximum-temperature series, the spring midpoint moves to earlier dates much more than the autumn midpoint, and the temperatures associated with the points increase by different amounts. These changes imply that the lengths of seasonal segments change, including a shortening of the cooler half of the year. This is one of the main strengths of the feature-based approach: timing and temperature can be studied together within a single set of point estimates. Five-year summaries further show that the emergence of earlier seasonal timing is clearest from the 2000s onward, even though warming at the seasonal points is consistently present across the full record. This distinction would be much less visible in a two-period comparison alone.

The present application focuses on establishing SALLI and illustrating its outputs using an Australia-wide maximum-temperature curve. This national summary reduces local variability and does not represent every climatic region or capture differences between maximum and minimum temperature. Our companion paper, currently in preparation, extends the analysis to both temperature measures across 15 countries and Australia's cool-winter, dry-summer and humid-summer groups. These extensions will help assess regional patterns that national aggregation may conceal and differences in seasonal timing and duration across temperature measures, climatic settings and hemispheres. This broader work could inform agricultural adaptation and advisory services \citep{psilou2026}, with attention to agricultural income and environmental outcomes \citep{zafeiriou2018}. Further assessment at individual stations, particularly in climates with weak or multiple annual temperature peaks, remains necessary to establish where the selected seasonal structure is appropriate.

\section{Conclusion}
\label{sec:conclusion}

We developed SALLI, a new multivariate data-driven index for defining and tracking thermal seasons from daily temperature data. SALLI characterises the annual temperature cycle through four unambiguous seasonal points: summer, autumn, winter, and spring and their fitted temperatures. It selects an optimal bandwidth \(h^\star\) for each yearly series, thereby allowing smoothing to adapt to the data's structure. In this way, SALLI provides a robust and transferable approach for seasonal timing analysis that captures both horizontal shifts in timing and changes in temperature within a single curve-based description. Applied to the Australia-wide daily maximum temperature series, SALLI highlights that recent decades are characterised by warming and an earlier seasonal cycle, with the clearest movement in spring and relatively little change in autumn. It also shows that these changes are not the same across the four seasonal points, implying changes in the lengths of seasonal segments as well as in the temperatures associated with them. More broadly, the Australian application demonstrates that SALLI not only recovers meaningful seasonal structure from daily data, but also reveals aspects of seasonal change, including unequal movement across seasonal points and the emergence of clearer earlier timing mainly after the 2000s, that are less accessible to threshold-based or fixed-smoother approaches.

\backmatter

\bibliography{sn-bibliography}

\section*{Statements and Declarations}

\bmhead{Author Contributions}
T.K.\ curated the data, performed the analyses, and wrote the first draft of the manuscript. I.K., G.B.\ and S.H.\ contributed to the study design, provided methodological advice, and revised the manuscript. All authors reviewed and approved the final manuscript.

\bmhead{Data and Code Availability}
The ACORN-SAT temperature data used in this study are publicly available from the Australian Bureau of Meteorology (\url{http://www.bom.gov.au/climate/data/acorn-sat/}). The processed data and analysis code used to generate the results are available from the corresponding author upon reasonable request.

\bmhead{Supplementary Information}
No supplementary materials are provided.

\bmhead{Funding}
This research was supported by an Australian Government Research Training Program (RTP) Stipend Scholarship administered through RMIT University. T.K.\ was also supported by a Natural Hazards Research Australia stipend top-up scholarship.

\bmhead{Competing Interests}
The authors declare that they have no competing interests.

\bmhead{Ethics Approval}
Not applicable.

\bmhead{Consent for Publication}
Not applicable.

\end{document}